\documentclass[11pt,twoside]{article}
\usepackage{amsmath,amssymb}
\numberwithin{equation}{section}

\begin{document}
\date{}
\title{\textbf{Stacked Central Configurations for Newtonian N+4-Body Problems }\footnote{This work is supported by NSF of China and Youth found of Mianyang Normal University.}}

\maketitle
\indent \indent \indent  \indent \indent \indent  \author{Furong Zhao$^{1,2}$\, and Shiqing Zhang$^1$ }

\begin{center}
$^1$Department of Mathematics, Sichuan University, Chengdu, 610064,P.R.China\\
$^{2}$Department of Mathematics and Computer Science, Mianyang Normal
University,
Mianyang, Sichuan,621000,P.R.China\\
\end{center}

\textbf{Abstract}:
In this paper,we study spatial central configurations where N  bodies are at the vertices of a regular N-gon $T$
and the other $4$ bodies are symmetrically located on the straight line that is perpendicular to the plane that contains $T$ and passes through the center of $T$.We study the necessary conditions about masses for the bodies which can form a central configuration and show the existence of central configurations for Newtonian N+4-body problems.
\baselineskip=20pt
\section{Introduction and Main Results }

\indent \indent The Newtonian n-body problems([1],[23])  concern with the motions
 of n particles with masses $m_j \in R^{+}$ and positions  $q_j \in R^{3}$$(j=1,2,... ,n)$ ,
 the motion  is governed by Newton's second law and the Universal law:\\
\begin{equation}
m_{j}\ddot{q}_{j}=\frac{\partial U(q)}{\partial {q}_{j}},
\end{equation} where $q=(q_{1},q_{2},\cdots,q_{n})$ and $U(q)$ is Newtonian potential:\\
\begin{equation}
U(q)=\sum_{1\leqslant j<k\leqslant n }\frac{m_{j}m_{k}}{|q_{j}-q_{k}|},
\end{equation}
Consider the space\\
\begin{equation}
X=\{q=(q_{1},q_{2},\cdots,q_{n})\in
R^{3n}:\sum_{j=1}^{n}m_{j}q_{j}=0 \},
\end{equation}
i.e,suppose that the center of mass is fixed at the origin of the
space. Because the potential is singular when two particles have
same position, it is natural to assume that the configuration
avoids the collision set $\triangle =\{q=(q_1,\cdots,q_n):q_j=q_k
$ for some $k\neq j\}$.The set $X\backslash \triangle $ is called
the configuration
space.\\
\indent \textbf{Definition 1.1}([20,24]):A configuration $q=(q_{1},q_{2},\cdots,q_{n})\in X\backslash \triangle$
is called a central configuration if there exists a constant $\lambda$ such that\\
\begin{equation}
\sum_{j=1,j\neq k}^{n}\frac{m_{j}m_{k}}{|q_{j}-q_{k}|^{3}}(q_{j}-q_{k})=-\lambda m_{k}q_{k},1\leqslant k\leqslant n.
\end{equation}
The value of constant $\lambda$ in (1.4) is uniquely determined by \\
\begin{equation}
\lambda=\frac{U}{I},
\end{equation}
Where
\begin{equation}
I=\sum_{k=1}^{n}m_{k}|q_{k}|^{2}.
\end{equation}
Since the general solution of the n-body problem can't be
given, great importance has been attached to search for particular solutions from the very beginning. A homographic solution is  a configuration which is preserved for all time. Central configurations and homographic solutions are linked by the Laplace theorem (see
[24]).Collaps orbits and parabolic orbits have relations with
the central configurations([17,19,20]).So finding central
configurations becomes very important. The main general open
problem for the cental configurations is due to Winter[24]and
Smale[22]:Is the number of  central configurations finite
for any choice of positive masses $m_1,...,m_n $?Hampton and Moeckel([6])
have proved this conjecture for four any  given positive masses. \\
For 5-body problem ,Hampton ([5])provided a new family of planar
central configurations,called stacked central configurations which
 has some proper subset of three or more points
forming a central configuration. \\
Ouyang ,Xie and Zhang([15]) studied pyramidal central configurations for Newtonian N+1-body problems; Zhang and Zhou([25]) studied double pyramidal central configurations for Newtonian N+2-body problems;
Mello and Fernandes([11]) studied new classes of spatial central configurations for the N+3-body problem.\\
Based the above works,we study stacked central configuration for Newtonian N+4-body problems.
 in N+4-body problems, for which  $N$ bodies are at the vertices the vertices of a  regular polygon ,
  the other $4$ bodies are symmetrically located on the straight line that is perpendicular to the plane that contains $T$ and passes through the center of $T$,the vertical line passes the geometrical center of the  regular polygon.(see Fig 1 for $N=4$).\\
 Related assumptions will be interpreted more precisely
in the following.\\
Without loss of generality we can take a coordinate system such that\\
$q_j=(\cos(\frac{(j-1)}{N}2\pi),\sin(\frac{(j-1)}{N}2\pi),0)$ where $j=1,\cdots,N$;\\
$q_{N+1}=(0,0,r_1)$,
$q_{N+2}=(0,0,-r_1)$,
$q_{N+3}=(0,0,r_2)$,
$q_{N+4}=(0,0,-r_2)$.\\
\begin{center}%8体中心构型
\setlength{\unitlength}{1mm}
\begin{picture}(80,40)
\thicklines
\put(15,20){\circle*{3}}
\put(15,23){$m_6$}
\put(5,20){\circle*{3}}
\put(5,23){$m_8$}

\put(0,20){\vector(1,0){80}}
\put(80,18){$z$}
\put(40,0){\vector(0,1){45}}
\put(41,43){$y$}
\multiput(22,8)(1.5,1){30}{\line(0,1){0.5}}
\put(49,26){\circle*{2}}
\put(52,26){$m_1$}
\multiput(40,40)(0.6,-0.9){17}{\line(0,1){0.2}}
\multiput(49,26)(-0.3,-0.9){28}{\line(0,1){0.2}}
\put(67,38){\vector(3,2){2}}
\put(67,35){$x$}
\put(40,40){\circle*{3}}
\put(43,40){$m_2$}
\put(40,40){\line(-1,-4){6}}
\put(34,16){\circle*{5}}
\put(31,10){$m_3$}

\put(40,0){\circle*{3}}
\put(43,0){$m_4$}
\multiput(40,40)(0.6,-0.9){17}{\line(0,1){0.2}}
\put(40,0){\line(-1,3){6}}

\put(65,23){$m_5$}
\put(65,20){\circle*{3}}

\put(75,23){$m_7$}
\put(75,20){\circle*{3}}
\put(10,0){Fig.1}
\end{picture}
\end{center}
We have :\\
\textbf{Theorem1.1}:If $m_N+1=m_N+2$(or $m_N+3=m_N+4$) and $m_1,\cdots,m_{N+4}$  form a central configuration,then \\
(1):$\Sigma_{j=1}^Nm_jq_j=0$\\
(2):$m_N+3=m_N+4$( $m_N+1=m_N+2$).\\
(3):$m_1,\cdots,m_{N}$ also form a central configuration.\\
(4):$m_1=\cdots=m_N$.\\
\textbf{Theorem1.2}:Assume that  $m_1=\cdots=m_N=1$,$m_{N+1}=m_{N+2}=M_1$,$m_{N+3}=m_{N+4}=M_1$,then there exist
$\epsilon(r_1,r_2)>0$ , $\delta>0$ such that $\forall$ $(r_1,r_2)$$\in$ $\{(r_1,r_2)|
 r_2>r_1>\delta , r_2-r_1<\epsilon(r_1,r_2)\}$,
 we have positive masses $M_1=M_1(r_1,r_2)$,$M_2=M_2(r_1,r_2)$ and all the
$N+4$ bodies form a central configuration.\\
\textbf{Remark 1}:$M_1=\frac{b_1a_{22}-b_2a_{12}}{a_{11}a_{22}-a_{12}a_{21}}$,
$M_2=\frac{b_2a_{11}-b_1a_{21}}{a_{11}a_{22}-a_{12}a_{21}}$.
\\Where :\\ $a_{11}=\frac{1}{4r_1^3}-\frac{2}{|1+r_1^2|^{3/2}}$,$a_{12}=\frac{1}{|r_1+r_2|^2r_1}-\frac{1}{|r_1-r_2|^2r_1}
-\frac{2}{|1+r_2^2|^{3/2}}$,\\
$a_{22}=\frac{1}{4r_2^3}-\frac{2}{|1+r_2^2|^{3/2}}$,
$a_{21}=\frac{1}{|r_1+r_2|^2r_2}+\frac{1}{|r_1-r_2|^2r_2}-\frac{2}{|1+r_1^2|^{3/2}}$.\\
$b_1=\lambda^*-\frac{N}{|1+r_1^2|^{3/2}}$\\
$b_2=\lambda^*-\frac{N}{|1+r_2^2|^{3/2}}$\\
\textbf{Remark 2}:When $N=2$ ,the \textbf{Theorem1.2} is related to the \textbf{Theorem 1.3} in [8].
\section{The Proofs of Theorems}
\subsection{Some Lemmas}
We need some Lemmas.\\
If $ n\times n$ matrix $A=(a_{ij})$ satisfies
\begin{equation}
a_{i,j}=a_{i-1,j-1},1\leq i,j\leq n ,
\end{equation}
where we assume $a_{i,0}=a_{i,n} , a_{0,j}=a_{n,j}$,then A is called a circulant matrix.\\
\textbf{Lemma2.1}(see [10]).Let $A=(a_{ij})$ be a circulant
matrix,then the eigenvalues $\lambda_k$ and eigenvectors
$\overrightarrow{v}_k$ of A are
\begin{equation}
\lambda_k(A)=\sum_{j=1}^na_{1,j}\rho_{k-1}^{j-1}
\end{equation}
and
\begin{equation}
\overrightarrow{v}_k=(\rho_{k-1},\rho_{k-1}^{2},\cdots,\rho_{k-1}^{n})^T
\end{equation}
where $\rho_{k}=e^{\sqrt{-1}\frac{2k\pi}{n}}$.\\
\textbf{Lemma2.2}([24]):For $n\geq 3$,and $m_{1}=m_{2}=\cdots=m_{n}$,
if $(m_{1},m_{2},\cdots,m_{n})$ locate at vertices of a regular polygon ,then they form a central configuration.\\
From (1.4)and (1.3),notice that we have
\begin{equation}
\begin{split}
\sum_{j=1,j\neq k}^{n}\frac{m_{j}m_{k}}{|q_{j}-q_{k}|^{3}}(q_{j}-q_{k})=-\lambda m_{k}q_{k}=-\lambda m_{k}(q_{k}-q_{0})\\=
-\lambda m_{k}(q_{k}-\frac{\sum_{j=1}^{n}m_jq_j}{M})
=-m_{k}\frac{\lambda}{M} \sum_{j=1}^{n}m_j(q_{k}-q_j)
\end{split}
\end{equation}
where $M=\sum_{j=1}^nm_j$,$q_{0}=\frac{\sum_{j=1}^{n}m_jq_j}{M}$,\\
So (1.4) is also  equivalent to
\begin{equation}
\sum_{j=1,j\neq k}^{n}m_{j}(\frac{1}{|q_{j}-q_{k}|^{3}}-\frac{\lambda}{M})(q_{j}-q_{k})=0,k=1,2,3,\cdots,n.
\end{equation}
\subsection{The Proofs of Theorem 1.1 and Theorem 1.2}
\subsubsection{The Proof of Theorem 1.1}

If $m_1,\cdots,m_{N+4}$ form a central configuration,we have
\begin{equation}
\sum_{j=1,j\neq
k}^{N+4}m_{j}(\frac{1}{|q_{j}-q_{k}|^{3}}-\frac{\lambda}{M})(q_{j}-q_{k})=0,
k=1,\cdots,N+4.
\end{equation}\\
Notice that (2.6) can be also written as :\\
\begin{equation}
\begin{split}
\sum_{j=1,j\neq k}^{N}m_{j}(\frac{1}{|q_{j}-q_{k}|^{3}}-\frac{\lambda}{M})(q_{j}-q_{k})+\\
\sum_{j=1}^{4}m_{N+j}(\frac{1}{|q_{N+j}-q_{k}|^{3}}-\frac{\lambda}{M})(q_{N+j}-q_{k})=0,
\\k=1,\cdots,N.
\end{split}
\end{equation}\\
and
\begin{equation}
\begin{split}
\sum_{j=1}^{N}m_{j}(\frac{1}{|q_{j}-q_{N+l}|^{3}}-\frac{\lambda}{M})(q_{j}-q_{N+l})+\\
\sum_{j=1,j\neq l}^{4}m_{N+j}(\frac{1}{|q_{N+j}-q_{N+l}|^{3}}-\frac{\lambda}{M})(q_{N+j}-q_{N+l})=0,
\\l=1,2,3,4.
\end{split}
\end{equation}\\
Now (2.8) is taken inner product with vectors $\overrightarrow{e}_1=(1,0,0)$
and $\overrightarrow{e}_2=(0,1,0)$,then we get:\\
\begin{equation}
\begin{split}
(\frac{1}{|q_{j}-q_{N+l}|^{3}}-\frac{\lambda}{M})\sum_{j=1}^{N}m_{j}\cos(\frac{(j-1)}{N}2\pi)=0\\
(\frac{1}{|q_{j}-q_{N+l}|^{3}}-\frac{\lambda}{M})\sum_{j=1}^{N}m_{j}\sin(\frac{(j-1)}{N}2\pi)=0\\
j=1,\cdots,N.
\end{split}
\end{equation}\\
(2.9) can be also written as
\begin{equation}
(\frac{1}{|q_{j}-q_{N+l}|^{3}}-\frac{\lambda}{M})\sum_{j=1}^{N}m_{j}q_j=0,j=1,\cdots,N.\\
\end{equation}
It is obvious that
\begin{equation}
(\frac{1}{|q_{j}-q_{N+l}|^{3}}-\frac{\lambda}{M})=(\frac{1}{|q_{k}-q_{N+l}|^{3}}-\frac{\lambda}{M}),1\leq k,j\leq N,
\end{equation}
we get\\
\begin{equation}
\sum_{j=1}^{N}m_{j}q_{j}=0
\end{equation}
(2.8) is taken inner product with vector $\overrightarrow{e}_3=(0,0,1)$,then we have:\\
\begin{equation}
\begin{split}
\sum_{j=1}^{N}m_{j}(\frac{1}{|1+r_1^2|^{3/2}}-\frac{\lambda}{M})r_1+
0m_{N+1}+
2r_1(\frac{1}{|2r_1|^3}-\frac{\lambda}{M})m_{N+2}+\\
(r_1-r_2)(\frac{1}{|r_1-r_2|^3}-\frac{\lambda}{M})m_{N+3}+
(r_1+r_2)(\frac{1}{|r_1+r_2|^3}-\frac{\lambda}{M})m_{N+4}=0
\end{split}
\end{equation}
\begin{equation}
\begin{split}
\sum_{j=1}^{N}m_{j}(\frac{1}{|1+r_1^2|^{3/2}}-\frac{\lambda}{M})r_1+
2r_1(\frac{1}{|2r_1|^3}-\frac{\lambda}{M})m_{N+1}+
0m_{N+2}+\\
(r_1+r_2)(\frac{1}{|r_1+r_2|^3}-\frac{\lambda}{M})m_{N+3}+
(r_1-r_2)(\frac{1}{|r_1-r_2|^3}-\frac{\lambda}{M})m_{N+4}=0
\end{split}
\end{equation}
\begin{equation}
\begin{split}
\sum_{j=1}^{N}m_{j}(\frac{1}{|1+r_2^2|^{3/2}}-\frac{\lambda}{M})r_2+
(r_2-r_1)(\frac{1}{|r_1-r_2|^3}-\frac{\lambda}{M})m_{N+1}+\\
(r_1+r_2)(\frac{1}{|r_1+r_2|^3}-\frac{\lambda}{M})m_{N+2}+
0m_{N+3}+
2r_2(\frac{1}{|2r_2|^3}-\frac{\lambda}{M})m_{N+4}=0
\end{split}
\end{equation}
\begin{equation}
\begin{split}
\sum_{j=1}^{N}m_{j}(\frac{1}{|1+r_2^2|^{3/2}}-\frac{\lambda}{M})r_2+
(r_1+r_2)(\frac{1}{|r_1+r_2|^3}-\frac{\lambda}{M})m_{N+1}+\\
(r_2-r_1)(\frac{1}{|r_1-r_2|^3}-\frac{\lambda}{M})m_{N+2}+
2r_2(\frac{1}{|2r_2|^3}-\frac{\lambda}{M})m_{N+3}+
0m_{N+4}=0
\end{split}
\end{equation}
By(2.13)and(2.14),we have:
\begin{equation}
\begin{split}
2r_1(\frac{1}{|2r_1|^3}-\frac{\lambda}{M}))(m_{N+1}-m_{N+2})+\\
[(r_1+r_2)(\frac{1}{|r_1+r_2|^3}-\frac{\lambda}{M})-
(r_1-r_2)(\frac{1}{|r_1-r_2|^3}-\frac{\lambda}{M})](m_{N+3}-m_{N+4})=0
\end{split}
\end{equation}
By(2.15)and(2.16),we have:
\begin{equation}
\begin{split}
[(r_2-r_1)(\frac{1}{|r_1-r_2|^3}-\frac{\lambda}{M})-
(r_1+r_2)(\frac{1}{|r_1+r_2|^3}-\frac{\lambda}{M})])(m_{N+1}-m_{N+2})+\\
2r_2(\frac{1}{|2r_2|^3}-\frac{\lambda}{M}))(m_{N+4}-m_{N+3})=0
\end{split}
\end{equation}
We define:$f(x)=x(\frac{1}{x^3}-\frac{\lambda}{M}))$,$\frac{df(x)}{dx}=-\frac{2}{x^3}-\frac{\lambda}{M}<0,$so
\begin{equation}
f(r_2-r_1)=(r_2-r_1)(\frac{1}{|r_1-r_2|^3}-\frac{\lambda}{M})\neq
(r_2+r_1)(\frac{1}{|r_1+r_2|^3}-\frac{\lambda}{M})=f(r_2+r_1)
\end{equation}
If $m_{N+1}=m_{N+2}$,by (2.17) and (2.18), we have $m_{N+3}=m_{N+4}$.\\
If $m_{N+3}=m_{N+4}$,by (2.17) and (2.18), we have $m_{N+1}=m_{N+2}$.\\
By $m_{N+1}=m_{N+2}$,$m_{N+3}=m_{N+4}$ and (2.7), we have
\begin{equation}
\sum_{j=1,j\neq k}^{N}m_{j}(\frac{1}{|q_{j}-q_{k}|^{3}}-\frac{\lambda}{M})(q_{j}-q_{k})=0,
k=1,\cdots,N.
\end{equation}
Since $q_j$ locates on a unit circle,let\\
$q_k=\exp(\frac{2(k-1)\pi i}{N})$,$i=\sqrt{-1}$.By (2.20) we have\\
\begin{equation}
\sum_{j=1,j\neq k}^{N}m_{j}(\frac{1}{|q_{j-k}-1|^{3}}-\frac{\lambda}{M})(q_{j-k}-1)=0,
k=1,\cdots,N.
\end{equation}
We define the $N\times N$ matrix $C=(c_{k,j})$,where\\
$c_{k,j}=0$,for $j=k$;$c_{k,j}=(\frac{1}{|q_{j-k}-1|^{3}}-\frac{\lambda}{M})(q_{j-k}-1)$,for $j\neq k$.\\
$C$ is circulant matrix since\\
$c_{k-1,j-1}=c_{k,j}=0$,for $j=k$;
$c_{k-1,j-1}=(\frac{1}{|q_{(j-1)-(k-1)}-1|^{3}}-\frac{\lambda}{M})(q_{(j-1)-(k-1)}-1)$
=$(\frac{1}{|q_{j-k}-1|^{3}}-\frac{\lambda}{M})(q_{j-k}-1)=c_{k,j}$for $j\neq k$.\\
Then (2.21) can be written as
\begin{equation}
CM^*=0
\end{equation}
where $M^*=(m_1,\cdots,m_N)^T$.\\
By \textbf{Lemma2.1} and (2.22) we have
\begin{equation}
 m_1=m_2=\cdots=m_N.
\end{equation}
By \textbf{Lemma2.2} and (2.23) we know that\\
$m_1,\cdots,m_{N}$ also form a central configuration.\\
The proof of \textbf{Theorem1.1} is completed.
\subsubsection{The Proof of Theorem 1.2}
Notice that $(q_1,\cdots,q_{N+4})$ is a central configuration if
and only if
\begin{equation}
\sum_{j=1,j\neq k}^{N+4}\frac{m_{j}m_{k}}{|q_{j}-q_{k}|^{3}}(q_{j}-q_{k})=-\lambda m_{k}q_{k},1\leqslant k\leqslant N+4.
\end{equation}
Since the symmetries,(2.24) is equivalent to
\begin{equation}
\sum_{j=1,j\neq k}^{N+4}\frac{m_{j}m_{k}}{|q_{j}-q_{k}|^{3}}(q_{j}-q_{k})=-\lambda m_{k}q_{k}, k=1,N+1,N+3.
\end{equation}
That is
\begin{equation}
\begin{split}
-\lambda(1,0,0)=-\lambda^*(1,0,0)+\frac{(-1,0,r_1)}{|1+r_1^2|^{3/2}}M_1
+\frac{(-1,0,-r_1)}{|1+r_1^2|^{3/2}}M_1\\
+\frac{(-1,0,r_2)}{|1+r_2^2|^{3/2}}M_2
+\frac{(-1,0,-r_2)}{|1+r_2^2|^{3/2}}M_2
\end{split}
\end{equation}
\begin{equation}
\begin{split}
-\lambda(0,0,-r_1)=\frac{N(0,0,r_1)}{|1+r_1^2|^{3/2}}+\frac{(0,0,2r_1)}{|2r_1|^{3}}M_1
+\frac{(0,0,r_1+r_2)}{|r_1+r_2|^{3}}M_2\\
+\frac{(0,0,r_1-r_2)}{|r_1-r_2|^{3}}M_2
\end{split}
\end{equation}
\begin{equation}
\begin{split}
-\lambda(0,0,-r_2)=\frac{N(0,0,r_2)}{|1+r_2^2|^{3/2}}
+\frac{(0,0,r_1+r_2)}{|r_1+r_2|^{3}}M_1
+\frac{(0,0,-r_1+r_2)}{|r_1-r_2|^{3}}M_1\\
+\frac{(0,0,2r_2)}{|2r_2|^{3}}M_2
\end{split}
\end{equation}
where $\lambda^*$ such that
\begin{equation*}
\sum_{j=1,j\neq k}^{N}\frac{m_{j}m_{k}}{|q_{j}-q_{k}|^{3}}(q_{j}-q_{k})=-\lambda^* m_{k}q_{k}, k=1,\cdots,N.
\end{equation*}
(2.26),(2.27)and (2.28) are equivalent to
\begin{equation}
\lambda=\lambda^*+\frac{2}{|1+r_1^2|^{3/2}}M_1
+\frac{2}{|1+r_2^2|^{3/2}}M_2
\end{equation}
\begin{equation}
\begin{split}
\lambda=\frac{N}{|1+r_1^2|^{3/2}}+\frac{1}{4r_1^3}M_1
+(\frac{1}{|r_1+r_2|^{2}r_1}
-\frac{1}{|r_1-r_2|^{2}r_1})M_2
\end{split}
\end{equation}
\begin{equation}
\begin{split}
\lambda=\frac{N}{|1+r_2^2|^{3/2}}
+(\frac{1}{|r_1+r_2|^{2}r_2}
+\frac{1}{|r_1-r_2|^{2}r_2})M_1
+\frac{1}{4r_2^{3}}M_2
\end{split}
\end{equation}
(2.29),(2.30)and (2.31) are equivalent to
\begin{equation}
\begin{split}
(\frac{1}{4r_1^3}-\frac{2}{|1+r_1^2|^{3/2}})M_1
+(\frac{1}{|r_1+r_2|^{2}r_1}
-\frac{1}{|r_1-r_2|^{2}r_1}-\frac{2}{|1+r_2^2|^{3/2}})M_2\\
=\lambda^*-\frac{N}{|1+r_1^2|^{3/2}}
\end{split}
\end{equation}
\begin{equation}
\begin{split}
(\frac{1}{|r_1+r_2|^{2}r_2}
+\frac{1}{|r_1-r_2|^{2}r_2}-\frac{2}{|1+r_1^2|^{3/2}})M_1
+(\frac{1}{4r_2^{3}}-\frac{2}{|1+r_2^2|^{3/2}})M_2\\
=\lambda^*-\frac{N}{|1+r_2^2|^{3/2}}
\end{split}
\end{equation}
when  $a_{11}a_{22}-a_{12}a_{21}\neq 0$,we have
\begin{equation}
M_1=\frac{b_1a_{22}-b_2a_{12}}{a_{11}a_{22}-a_{12}a_{21}}
\end{equation}
\begin{equation}
M_2=\frac{b_2a_{11}-b_1a_{21}}{a_{11}a_{22}-a_{12}a_{21}}
\end{equation}
Where :\\ $a_{11}=\frac{1}{4r_1^3}-\frac{2}{|1+r_1^2|^{3/2}}$,$a_{12}=\frac{1}{|r_1+r_2|^2r_1}-\frac{1}{|r_1-r_2|^2r_1}
-\frac{2}{|1+r_2^2|^{3/2}}$,\\
$a_{22}=\frac{1}{4r_2^3}-\frac{2}{|1+r_2^2|^{3/2}}$,
$a_{21}=\frac{1}{|r_1+r_2|^2r_2}+\frac{1}{|r_1-r_2|^2r_2}-\frac{2}{|1+r_1^2|^{3/2}}$.\\
$b_1=\lambda^*-\frac{N}{|1+r_1^2|^{3/2}}$\\
$b_2=\lambda^*-\frac{N}{|1+r_2^2|^{3/2}}$\\
If \begin{equation}
a_{11}a_{22}-a_{12}a_{21}<0,b_1a_{22}-b_2a_{12}<0,b_2a_{11}-b_1a_{21}<0,
\end{equation}
then
\begin{equation}
M_1>0,M_2>0
\end{equation}
Notice that (2.36) is equivalent to
\begin{equation}
\frac{a_{11}}{a_{21}}<\frac{b_1}{b_2}<\frac{a_{12}}{a_{22}}
\end{equation}
Notice that
\begin{equation}
\begin{split}
\frac{a_{11}}{a_{21}}=\frac{\frac{1}{4r_1^3}-\frac{2}{|1+r_1^2|^{3/2}}}{\frac{1}{|r_1+r_2|^2r_2}+\frac{1}{|r_1-r_2|^2r_2}-\frac{2}{|1+r_1^2|^{3/2}}}\\
=\frac{|1+r_1^2|^{3/2}-8r_1^3}{4r_1^3\times |1+r_1^2|^{3/2}}\times
\frac{(r_2^2-r_1^2)^2r_2(1+r_1^2)^{3/2}}{2(r_1^2+r_2^2)(1+r_1^2)^{3/2}-2(r_2^2-r_1^2)^2}\\
=\frac{|1+r_1^2|^{3/2}-8r_1^3}{4r_1^3}\times
\frac{(r_2^2-r_1^2)^2r_2}{2(r_1^2+r_2^2)(1+r_1^2)^{3/2}-2(r_2^2-r_1^2)^2r_2}\\
\end{split}
\end{equation}
Since $
\lim_{r_1\rightarrow+\infty}\frac{|1+r_1^2|^{3/2}-8r_1^3}{4r_1^3}=-\infty,$
and for $r_2=r_1$,$2(r_2^2-r_1^2)^2 r_2=0$. Then there exists
$\delta_1>0$,$\epsilon(r_1,r_2)>0$,such that for $r_1>\delta_1,$
we have
$$\frac{|1+r_1^2|^{3/2}-8r_1^3}{4r_1^3}<0,$$
and for $r_2-r_1<\epsilon(r_1,r_2),$ we have
$$2(r_1^2+r_2^2)(1+r_1^2)^{3/2}-2(r_2^2-r_1^2)^2r_2>0$$
So
\begin{equation}
\begin{split}
\frac{a_{11}}{a_{21}}<0,\forall (r_1,r_2)\in \{(r_1,r_2)|
r_2>r_1>\delta , r_2-r_1<\epsilon(r_1,r_2)\}
\end{split}
\end{equation}
We also notice that
\begin{equation}
\lim_{r_1\rightarrow+\infty}\frac{b_1}{b_2}=1
\end{equation}
\begin{equation*}
\begin{split}
\frac{a_{12}}{a_{22}}=\frac{4r_2(1+r_2^2)^{3/2}+2(r_2^2-r_1^2)^2}{(r_2^2-r_1^2)^2(1+r_2^2)^{3/2}}
\times \frac{4r_2^3(1+r_2^2)^{3/2}}{8r_2^3-(1+r_2^3)^{3/2}}\\
=\frac{16(1+\frac{1}{r_2^2})^{3/2}+8(1-(\frac{r_1}{r_2})^2)}{(1-(\frac{r_1}{r_2})^2)(8-(1+\frac{1}{r_2^2})^{3/2})}
\end{split}
\end{equation*}
There exists $\delta_2>0$ ,such that for $r_2>\delta_2>0,$ we have
\begin{equation}
\frac{a_{12}}{a_{22}}>2
\end{equation}
By (2.40),(2.41)and(2.42), there exists
 $\delta\geq\max{\{\delta_1,\delta_2\}}$,such that for $\delta<r_1<r_2$ and $r_2-r_1<\epsilon(r_1,r_2),$
 we have
 $$\frac{a_{11}}{a_{21}}<\frac{b_1}{b_2}<\frac{a_{12}}{a_{22}}$$

The proof of \textbf{Theorem1.2} is completed.

\end{document}